%% file: main.tex
\documentclass{article} 
\usepackage{iclr2024_conference,times}

\input{math_commands.tex}

\usepackage{hyperref}
\usepackage{url}
\usepackage{graphicx}
\usepackage{natbib}

\title{Building Ocean Climate Emulators   }

\author{Adam Subel \& Laure Zanna   \thanks{Correspondence: adam.subel@nyu.edu} \\
Courant Institute of Mathematical Sciences\\
New York University
}

\iclrfinalcopy 
\begin{document}

\maketitle

\begin{abstract}

The current explosion in machine learning for climate has led to skilled, computationally cheap emulators for the atmosphere. However, the research for ocean emulators remains nascent despite the large potential for accelerating coupled climate simulations and improving ocean forecasts on all timescales. There are several fundamental questions to address that can facilitate the creation of ocean emulators. Here we focus on two questions: 1) the role of the atmosphere in improving the extended skill of the emulator and 2) the representation of variables with distinct timescales (e.g., velocity and temperature) in the design of any emulator.  In tackling these questions, we show stable prediction of surface fields for over 8 years, training and testing on data from a high-resolution coupled climate model, using results from four regions of the globe. Our work lays out a set of physically motivated guidelines for building ocean climate emulators.

\end{abstract}

\section{Introduction}
In the last few years, we have seen substantial growth in the development of machine-learning (ML) methods applied to weather and climate problems. In particular, many have successfully built data-driven emulators to provide low-cost representations for the complex dynamics of the climate system \citep{beucler2023machine}. Several emulators have been shown to perform as well or better than state-of-the-art numerical weather models on short-time scales \citep{bi2023accurate,kochkov2023neural,pathak2022fourcastnet}.

Beyond weather, the related problem of long-term atmosphere emulation, that is, time scales spanning ten to hundreds of years, has emerged in the literature \citep{kochkov2023neural,bonev2023spherical,watt2023ace}.
Recently, three applications of ocean emulation have shown some success: an idealized model for multi-decadal timescales \citep{bire2023ocean}, a regional emulation on subseasonal timescales \citep{chattopadhyay2023oceannet}, and a 3D global emulation on 30-day timescales \citep{xiong2023ai}. The emulation of ocean dynamics involves a different set of challenges to that of the atmosphere. As we continue to confront the growing risks of a changing climate, the need to develop reliable models to simulate the global climate will prove essential for adaptation and mitigation.

In this work, we highlight key considerations for building emulators of regional (and global) ocean climate models by assessing the role of boundary conditions and designing a framework for multi-scale emulation (e.g., different spatio-temporal scales).  Understanding the appropriate boundary inputs and scales on which to model velocity and temperature is critical for successfully emulating complex multi-scale ocean physics. Using a simple network design, a U-Net \citep{ronneberger2015u}, we can run a large array of experiments that explore the sensitivities of the model to different inputs and time steps. Our results, which focus on emulating surface ocean fields, in which multi-scale physics is a determining factor, should help guide future implementations of more costly, state-of-the-art machine learning techniques.

\section{Methods}
\subsection{Data}
We use data from the last 20 years of a pre-industrial control simulation of the GFDL CM2.6 coupled climate model, with a nominal ocean resolution of $1/10^\circ$ \citep{delworth2012simulated}. We focus on emulating four regions (the Gulf Stream, the Tropics, the South Pacific, and the African Cape, see fig.~\ref{fig:Region_Figure}), which provide a range of dynamics for evaluating the emulators. The data are regridded from the native ocean resolution of $1/10^\circ$ onto a grid with a resolution of $1/4^\circ$ (equivalent to a coarser version of the same climate model, CM2.5). To remove mesoscale turbulence, a source of short-term, local variability, the data is further filtered at a fixed factor corresponding to $1^\circ$ \citep{Loose2022}.  For the atmospheric data, which is natively on a $1/2^\circ$-resolution grid, we use bilinear interpolation of the data directly onto the $1/4^\circ$-resolution grid.

We will consider the state vector predicted by the network to be $\boldsymbol{\Phi}=(u,v,T)$, which represents the zonal velocity, meridional velocity, and temperature, respectively, in the surface layer. We let $\boldsymbol{\tau} = (\tau_u,\tau_v,T_{\mathrm{atm}})$ represent the state vector for the surface atmosphere boundary conditions, which are zonal wind stress, meridional wind stress, and air temperature respectively. We include details on the training and testing splits in the appendix.

\subsection{Input Features}
\label{sec:Lateral_Boundaries}
Our network is a U-Net with four upsampling and downsampling layers, yielding $\mathcal{O}(10^7)$ parameters. It takes in an input vector with 9 channels: three channels correspond to the ocean state $\boldsymbol{\Phi}=(u,v,T)$, three channels for the atmospheric boundary $\boldsymbol{\tau} = (\tau_u,\tau_v,T_{\mathrm{atm}})$, and another three channels for the lateral boundary conditions of $\boldsymbol{\Phi}$. 

To construct our regional emulators, we follow the standard practice in regional ocean modeling and using  the true state at the open boundaries to capture inflow and outflow dynamics correctly \citep{marchesiello2001open}. To encourage the network to learn a specific relationship between the open boundaries and the interior cells, we provide input channels of the boundary halo taken to be 4 cells wide. The standard input channels are kept as the total size of both the interior and the halo, and the network outputs the full field, including the halo region. At the time of testing, to avoid divergence between the interior and the true boundary condition, the halo region of the prediction is replaced with the true boundary values at each time step (see figure \ref{fig:lat_diagram} for a diagram of this process).

\subsection{Training}
For training the network, we perform multi-step predictions to create a loss function that captures dynamics beyond the time step of the emulator, $\Delta t$. For convenience, we use the following notation for recurrent passes of the network: $\tilde{\boldsymbol{\Phi}}_{t+n\Delta t} = F_\theta^{(n)}(\boldsymbol{\Phi}_t,\boldsymbol{\tau}_t)$, where $^{(n)}$ indicates the number of recurrent passes, $\tilde{\cdot}$ is a predicted state, and $F_\theta$ is the neural network.
The loss function is a combination of mean squared error with an additional point-wise loss on kinetic energy such that:
\begin{equation}
     \mathcal{L}^{(N)} = \sum_{n=1}^{N}{\lambda_n\left(\left(1-\alpha \right)\mathcal{L}_{\mathrm{mse}}^{(n)} + \alpha\mathcal{L}_{\mathrm{KE}}^{(n)}\right)}.
 \end{equation}
\noindent Here, $\mathcal{L}$ is the total loss function, $\mathcal{L}_{\mathrm{mse}}= \left \Vert \boldsymbol{\Phi}_{t+n\Delta t} - F_\theta^{(n)}(\boldsymbol{\Phi}_t,\boldsymbol{\tau}_t) \right \Vert_2^2$ is the MSE loss and $\mathcal{L}_{\mathrm{KE}}= \left \Vert \left(u_{t+n\Delta t}^2 +v_{t+n\Delta t}^2 \right) - \left(\tilde{u}_{t+n\Delta t}^2 +\tilde{v}_{t+n\Delta t}^2 \right)\right \Vert_2^2$ is the loss contribution from kinetic energy. $N$ is the total number of recurrent passes. $\lambda_n$ are parameters weighting the loss after each recurrent pass; these are weighted highest for $n$ close to $N$. $\alpha$ is a balancing parameter to weight between the MSE loss and the kinetic energy loss, we choose $\alpha = .05$. To avoid over-complicating the loss from the start, the value of $N$ is increased incrementally as the network converges for each $N$. Unless otherwise specified, $N=4$ and $\Delta t =1$ day.

\section{Results}

\subsection{Impact of Boundary Information}

Accounting for the role of the atmosphere as an external forcing is essential for emulating the evolution of the ocean, especially for surface fields.
Providing inputs with a seasonal imprint is necessary when using a training window, $N \times\Delta t$, much shorter than seasonal timescales. For example, ocean SSTs are beneficial for building long-term atmosphere emulators \citep{watt2023ace}. Here, we explore the role of wind stress and surface air temperature in increasing the performance of the ocean emulator.

\begin{figure}[h]
    \centering
    \includegraphics[width=0.7\linewidth]{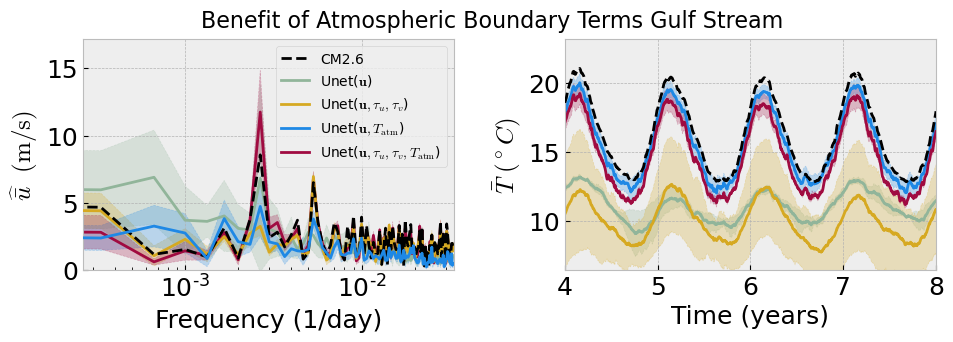}
    \caption{Impact of atmospheric boundary information, showing the skill of each emulator in performing long roll-outs (prediction). Left: the time spectra of the mean zonal velocity; Right: mean temperature timeseries. The shading shows the standard deviation across rollouts for 3 initial conditions.}
    \label{fig:Vary_Boundary_Main}
\end{figure}

In Fig.~\ref{fig:Vary_Boundary_Main}, the emulator provided with no atmospheric boundary information (green line) drifts in all predicted fields. Providing the wind stresses (yellow line) helps the emulator recover most of the missing dynamics in velocity, but it deteriorates the seasonal cycle; the skill in temperature prediction is only marginally changed. The addition of surface air temperature (blue line) has the most pronounced benefit due to its high decorrelation time; however, this emulator fails to capture the full range of time scales in velocity. Providing both boundary inputs (red line) leads to the best balance of performance across fields. It suggests that both high and low-frequency boundary information are necessary for emulating the surface ocean.

Since regional ocean models are constructed with open boundaries,

we build in the lateral information yields, which yield only small performance gains, even with the full set of $\tau_u$, $\tau_v$, and $T_{atm}$ passed to the model (see appendix).

\subsection{Capturing Multi-timescale dynamics}
\begin{figure}[ht]
    \centering
    \includegraphics[width=0.75\linewidth]{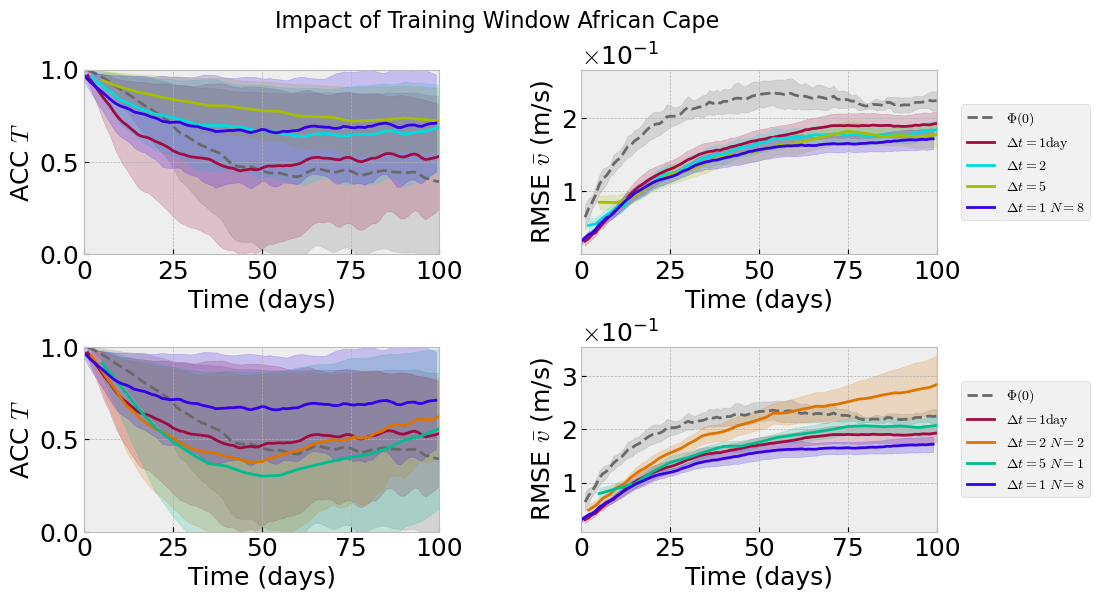}
    \caption{Impact of time step and effective training window on the short-term skill of the network for the African Cape region. ACC for temperature (left) and RMSE for meridional velocity (right). The top panels showcase the benefit of increasing $\Delta t$, and the bottom panels demonstrate the performance loss when decreasing the number of rollouts, $N$. The shading indicates the standard deviation across rollouts from 5 initial conditions and using 3 networks trained from different initial weights.}
    \label{fig:Lag_Main}
\end{figure}

\begin{figure}[ht]
    \centering
    \includegraphics[width=0.8\linewidth]{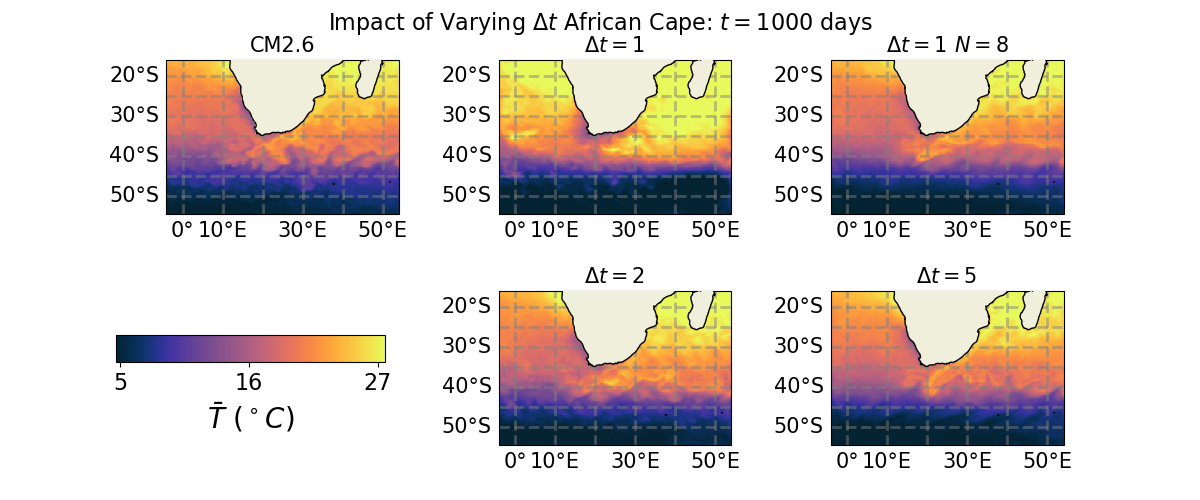}
    \caption{Snapshots of temperature shown for a 1000 days prediction, generated by the emulators with different $\Delta t$ and $N$ (as in fig.~\ref{fig:Lag_Main}.) }
    \label{fig:snap_main}
\end{figure}

Previous studies have investigated, or at least suggested, that the training window of the model can impact performance \citep{bire2023ocean,bi2023accurate}. Here we address this question for ocean surface emulation evolving at multiple timescales. We perform a sensitivity study to identify the necessary conditions for improved skill. Our baseline emulator, with $\Delta t=1$ and $N=4$, captures the velocity dynamics well but struggles to emulate the slowly evolving field, temperature. Figure \ref{fig:Lag_Main} demonstrates this as the anomaly correlation coefficient (ACC) of temperature drops quickly for the baseline emulator (red). This error accumulates and drives a drift for the temperature in the African Cape region (see appendix for long rollouts results). Increasing $\Delta t$ leads to an improvement in the ACC for temperature, with an increase in skill in 10-day ACC from 0.73 to 0.84 at $\Delta t = 2$ (cyan) and a further increase to 0.91 with $\Delta t = 5$ (green); see temperature snapshots in fig.~\ref{fig:snap_main}.

We compare the benefits of increasing the time step against those of increasing the number of recurrent passes used to train our baseline emulator, matching the training window of the network with $\Delta t = 2$. This emulator, with $N =8$ (blue line; ACC at 10 days of 0.82), shows a similar skill in predicting temperature to the emulator with $\Delta t = 2$, and the best skill across all velocity metrics. Reducing the training window for an emulator is detrimental to skill, as shown in the bottom panels of fig.~ \ref{fig:Lag_Main}. Networks with longer time steps but with a similar training window to the baseline perform poorly and sometimes become unstable. Capturing sufficiently long dynamics during training is necessary for an emulator to perform well but is a challenge given the slow timescales in the ocean (as compared to the atmosphere).

We note that regions with low variance in temperature, such as the tropics, do not show the same need for increased $\Delta t$ or $N$. Further details in the appendix show preliminary emulation results for ocean fields at a depth of $525~ \mathrm{m}$ and indicate a similar need for long training windows.

\section{Conclusion and Future Work}
Building computationally inexpensive ocean emulators for multi-decadal time scales can be beneficial for assessing the impacts of climate change, in particular, generating large ensembles for better estimates of uncertainty. This work identifies drivers of model skill and lays a foundation for the future construction of long-term ocean emulators. We demonstrate the role boundary conditions play in ocean emulation. We highlight the importance of surface air temperature for recovering long-time scales and wind stress for capturing shorter-time scale dynamics. We show that emulators need to use training windows spanning several days to accurately capture the evolution of slowly evolving fields with high variance, in this case, temperature. This can be achieved by increasing the time step or the number of recurrent passes during training.

By limiting our tests to the ML-model agnostic features of the emulator, we intend for our results to hold across a range of architectures and emulation tasks. We chose an ML architecture that is cheap and flexible (.013 seconds per simulated day; see appendix for performance details) but not the most advanced, and in some regions, we end up with under-energized flows at long prediction horizons. We will explore state-of-the-art ML methods, such as neural operators or transformers, following the principles developed in this work. We also intend to examine full 3D emulation (rather than each layer independently), where capturing a broader continuum of time scales and interaction between layers may pose additional challenges.

\subsubsection*{Acknowledgments}
We thank Alistair Adcroft and Chris Pedersen for many insightful conversations as well as the M$^2$LInES team for feedback and discussions. We acknowledge NOAA and GFDL for the model data used to perform experiments. This material is based upon work supported by the National Science Foundation Graduate Research Fellowship under Grant No. (DGE-2234660). This project is supported by Schmidt Sciences.

\bibliography{main}
\bibliographystyle{iclr2024_conference}

\appendix
\section*{Appendix}

\subsection*{Additional Information on Methods}

\subsubsection*{Data for Training and Testing}
\begin{figure}[h]

    \centering
    \includegraphics[width=.75\linewidth]{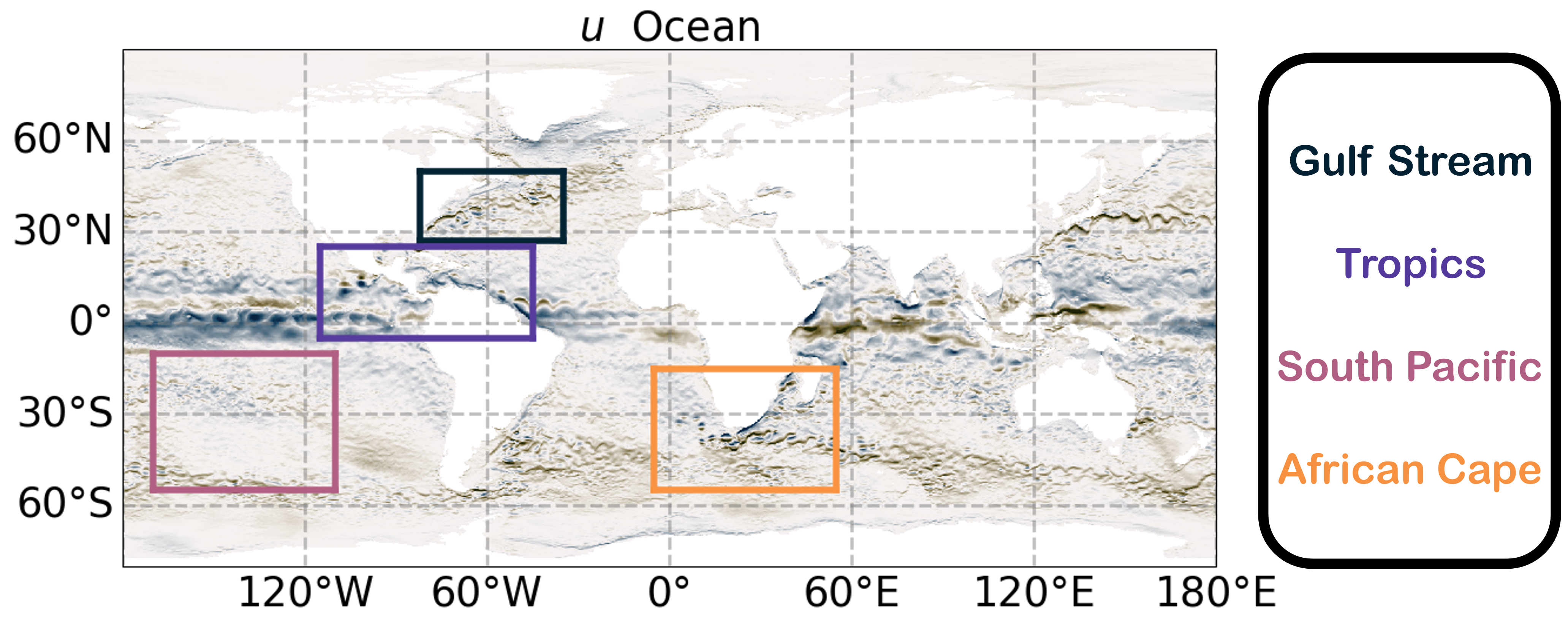}
    \caption{Global ocean surface zonal velocity, and 4 domains considered.}
    \label{fig:Region_Figure}
\end{figure}

We include outlines of the regions considered in figure \ref{fig:Region_Figure}. For surface emulation, the 20 years of data corresponds to 7305 samples using a calendar that includes leap years. We train on the first 4000 days of the model, saving 200 days for our validation dataset. For testing the model on long rollouts, we select our initial condition to be 200 days after the training dataset and run our model for 3000 days,  or 8.21 years. We compute our uncertainty ranges over runs performed by 3 emulators, each with a distinct random seed for initializing weights and training. In figure \ref{fig:Vary_Boundary_Main} we also include two additional runs of 3000 days from initial conditions 100 and 250 days after the training set. For testing the performance on short rollouts, we choose 5 initial conditions, starting 200 days after training. Each subsequent initial condition is then selected 200 days after the previous so that we sample from different points in the seasonal cycle. We compute our uncertainty over these 5 initial conditions using the same three networks mentioned above, yielding a total of 15 rollouts to compare.

For the subsurface fields, the 20 years of data corresponds to 1459 samples. We train on the first 1000 data samples and use the next 50 samples as validation. The last 400 samples, 2000 days, are held for testing.

\subsubsection*{Computing Wind Stress}
To compute the components of wind stress, we use the values of 10m wind velocities and wind speed along with a constant value for density ($\rho = 1.293\times 10 ^{-3} \mathrm{\frac{kg}{m^3}}$) and drag coefficient ($C_d = 1.2\times 10 ^{-3}$). Though the drag coefficient may vary in both space and time, we simplify the estimation of wind stress by choosing the constant in a reasonable range of expected values  \citep{kara2007wind}.

\subsubsection*{Architecture}
We use the same U-net architecture for all networks in this paper. Each network has $\mathcal{O}(1e7)$ parameters. The network is built on 4 downsampling and upsampling operations, with the number of channels going from $N_{in}$ to 64 to 128 to 256 to 512. Here, $N_{in}$ is equal to the number of channels in $\boldsymbol{\Phi}$, $\boldsymbol{\tau}$, and $\boldsymbol{\Phi}_{\mathrm{lateral}}$. There is a batch norm operation between each convolutional layer, except for the final output layer.

\subsubsection*{Computational Cost}
The model we use here is lightweight, and we can train models quickly. Using two Nvidia RTX8000 GPUs the model trains in an hour or less. For cases where we train on more than 4 rollout steps, we run the model on four RTX8000s and train in less than 3 hours. To give a more exact sense of the runtime cost of the model, we provide table \ref{tab:cost table} showing the cost of running a leap year (366 days) of simulation for each model on a single RTX8000 GPU. In the largest, and thus slowest, region, this corresponds to a runtime of 0.013 seconds per day.
\begin{table}[h]
    \centering
    \begin{tabular}{|c|c|c|} \hline
         Region&  Domain Size ($N_x$, $N_y$)&  Simulation Cost per Year \\ \hline
         South Pacific&  $224 \times 241$&  $4.72 \pm 0.07$ Seconds\\ \hline
         Gulf Stream&  $119\times 189$&  $2.34 \pm 0.04$ Seconds\\ \hline
         African Cape&  $204 \times 241$&  $4.29 \pm 0.06$ Seconds\\ \hline
         Tropics&  $124\times 281$&  $3.31 \pm 0.04$ Seconds\\ \hline
    \end{tabular}
    \caption{Relative computational cost between regions of running one simulated year using a single Nvidia RTX8000 GPU.}
    \label{tab:cost table}
\end{table}
\begin{figure}
    \centering
    \includegraphics[width=0.75\linewidth]{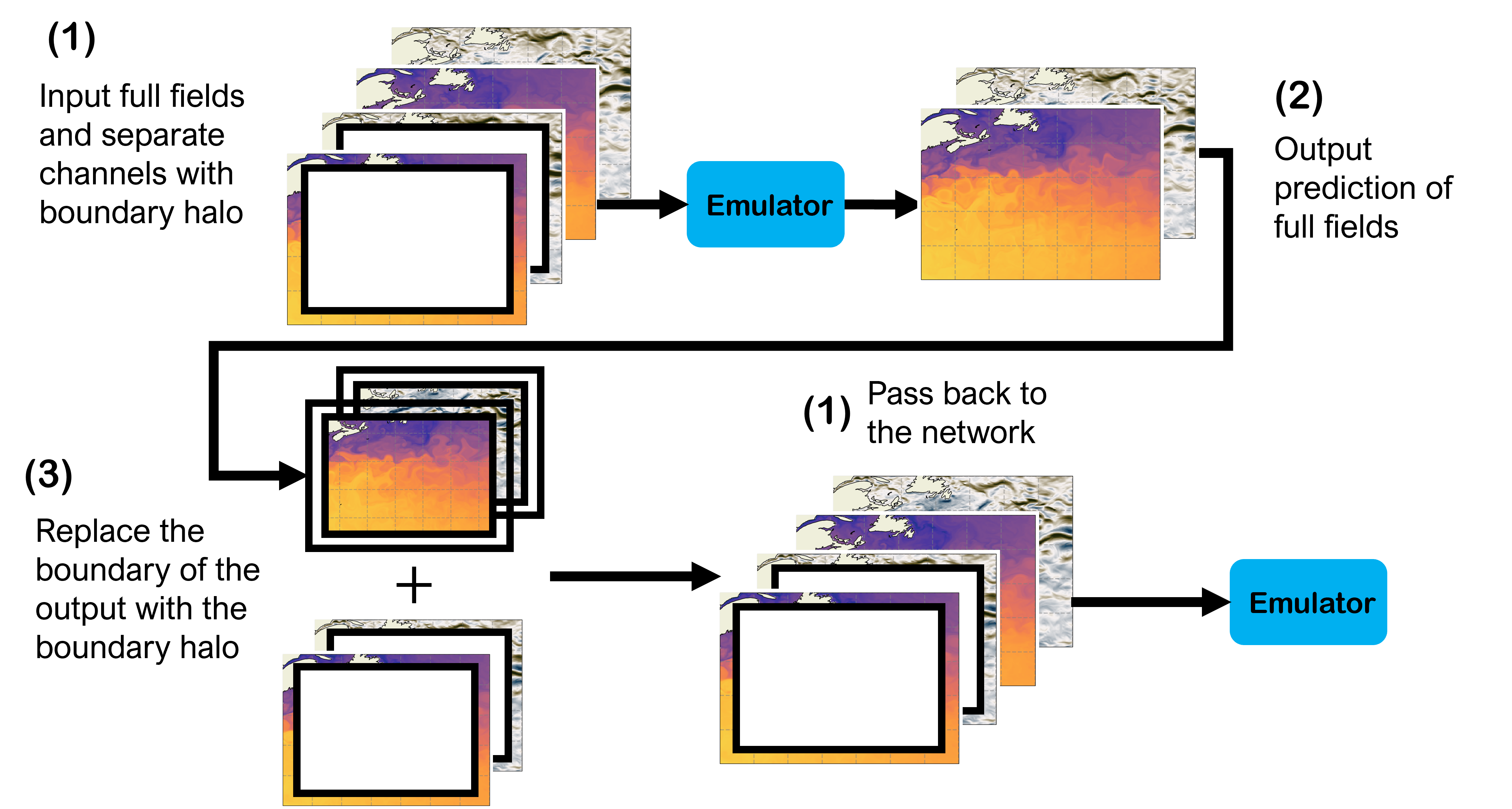}
    \caption{Methodology used to incorporate lateral boundary conditions into the model (section \ref{sec:Lateral_Boundaries}).}
    \label{fig:lat_diagram}
\end{figure}

\subsection*{Metrics}
To evaluate the performance of our network, we define a series of metrics that evaluate the emulator on short- and long-timescale predictions.

\subsubsection*{Short-Time Skill}

To quantify the performance over short time scales, we look at metrics of correlation coefficient (CC), anomaly correlation coefficient (ACC), and root mean squared error (RMSE). These take the following forms:

\begin{eqnarray}
    CC(u,\tilde{u}) = \frac{\sum_{i,j}^{N_x,N_y}A_{i,j}W_{i,j}u_{i,j}\tilde{u}_{i,j}}{\sqrt{\sum_{i,j}^{N_x,N_y}\left( A_{i,j}W_{i,j}u_{i,j} \right)^2\sum_{i,j}^{N_x,N_y}\left( A_{i,j}W_{i,j}\tilde{u}_{i,j} \right)^2}} \\
    ACC(u,\tilde{u}) = \frac{\sum_{i,j}^{N_x,N_y}A_{i,j}W_{i,j}(u_{i,j}-C_{i,j})(\tilde{u}_{i,j}-C_{i,j})}{\sqrt{\sum_{i,j}^{N_x,N_y}A_{i,j}W_{i,j}\left( u_{i,j}-C_{i,j} \right)^2\sum_{i,j}^{N_x,N_y}A_{i,j}W_{i,j}\left( \tilde{u}_{i,j}-C_{i,j} \right)^2}}\\
    RMSE(u,\tilde{u}) = \sqrt{\frac{\sum_{i,j}^{N_x,N_y}A_{i,j}W_{i,j}(u_{i,j}-\tilde{u}_{i,j})^2}{\sum_{i,j}^{N_x,N_y}A_{i,j}W_{i,j}}}
\end{eqnarray}

\noindent In these equations, $N_x,N_y$ are the number of grid points in the zonal and meridional directions respectively. $A_{i,j}$ is the cell area for the point $i,j$. $W_{i,j}$ is the wet mask at the point, 1 if the point is ocean and 0 if the point is land. $C_{i,j}$ is the pointwise climatology for the day of year computed over the entire 20 year dataset.

\subsubsection*{Long-Time Skill}
To quantify the skill of the model at longer time scales, we look at how well it captures the mean trends of the system. The primary metrics are capturing the area-weighted means and variances of the system as well as comparing the time-spectrum of the means. The formulas for computing the mean and variance are simply:

\begin{eqnarray}
    \overline{u}(t) = \frac{\sum_{i,j}^{N_x,N_y}A_{i,j}W_{i,j}u_{i,j}}{\sum_{i,j}^{N_x,N_y}A_{i,j}W_{i,j}}\\
    \mathrm{Var}({u}(t)) = \frac{\sum_{i,j}^{N_x,N_y}A_{i,j}W_{i,j}(u_{i,j}-\overline{u}(t))^2}{\sum_{i,j}^{N_x,N_y}A_{i,j}W_{i,j}}
\end{eqnarray}

\subsection*{Impact of Lateral Boundaries}
\begin{figure}
    \centering
    \includegraphics[width=1\linewidth]{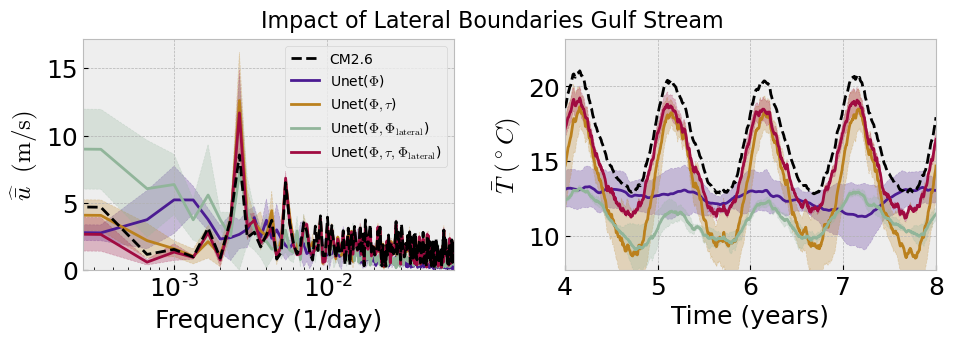}
    \caption{Impact of lateral boundary information, showing the skill of each emulator in performing long roll-outs. Left: the time spectra of the mean zonal velocity; Right: mean temperature timeseries. The shading shows the standard deviation across rollouts from 3 networks trained from different initial weights. }
    \label{fig:lateral_Gulf}
\end{figure}

In this section we show additional results demonstrating the benefit of including lateral boundary conditions. For the lateral boundary plots we omit the results from the African Cape region as all models perform poorly without a longer training interval or time step. Fig \ref{fig:lateral_Gulf} shows a case where the lateral boundary conditions yield a significant improvement in predicting temperature. The distance between the orange (without lateral boundaries) and red line shows that better capturing the open boundaries prevent the emulator from producing too cold winters. Also, noteworthy across all regions is that the model with no boundary conditions (purple line), lateral or from the atmosphere, produces no seasonal cycle in any field. In the South Pacific, a quiescent region with less fluid advected in and out of the domain, adding lateral boundaries does not lead to significant improvement when the emulator has access to the atmospheric boundary conditions. In these plots $\boldsymbol{\Phi}_{\mathrm{lateral}}$ references the addition of the lateral boundary halos as inputs.
\begin{figure}
    \centering
    \includegraphics[width=1\linewidth]{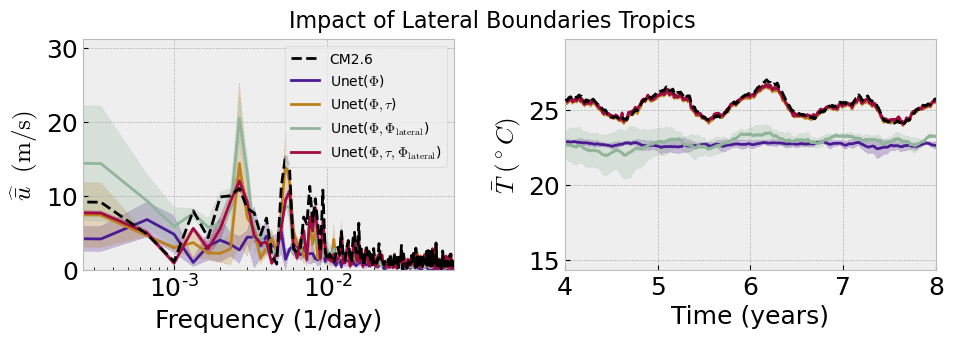}
    \caption{Impact of lateral boundary information, showing the skill of each emulator in performing long roll-outs. Left: the time spectra of the mean zonal velocity; Right: mean temperature timeseries. The shading shows the standard deviation across rollouts from 3 networks trained from different initial weights.}
    \label{fig:enter-label}
\end{figure}
\begin{figure}
    \centering
    \includegraphics[width=1\linewidth]{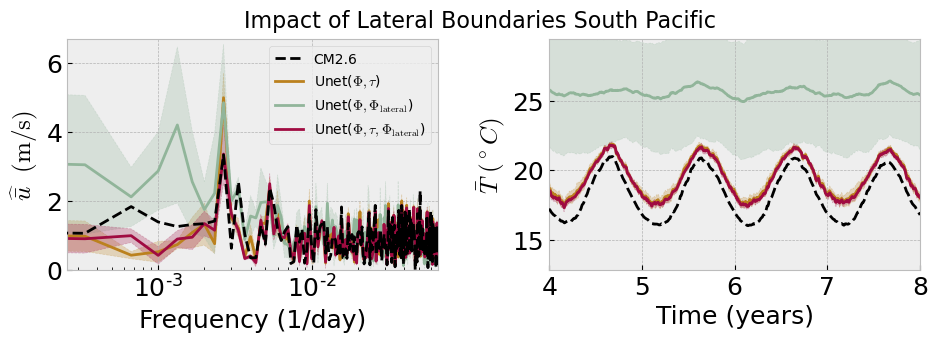}
    \caption{Impact of lateral boundary information, showing the skill of each emulator in performing long roll-outs. Left: the time spectra of the mean zonal velocity; Right: mean temperature timeseries. The shading shows the standard deviation across rollouts from 3 networks trained from different initial weights. The case for the U-Net with no surface or lateral boundary input is not shown as it becomes completely unstable.}
    \label{fig:enter-label}
\end{figure}

\subsection*{Subsurface Experiments}
As a preliminary exploration to help guide future work in emulation that resolves the vertical structure of the ocean, we performed a set of experiments in the same configuration as the main text but with different $\boldsymbol{\Phi}$ and $\boldsymbol{\tau}$. In these experiments we select an ocean depth below the mixed layer, $z=525 ~ \mathrm{m}$, where the influence of the atmosphere is near zero and define our $\boldsymbol{\Phi} = (u_{525},v_{525},T_{525})$ and our boundary conditions from the ocean surface as $\tau = (u_{\mathrm{surf}},v_{\mathrm{surf}},T_{\mathrm{surf}})$.

For the ocean depth, we have access to the 20 years of data; however, the fields are stored as 5-day averages, thus reducing the amount of data available for training and testing by a factor of 5. This reduction in data, alongside the longer timescales at depth, leads to poorer performance of the models and some drift in the rollouts. Despite this, we obtained meaningful results, even with the lower skill of the models, by comparing the results across time scales. Similar to the rest of the paper, we find that boundary conditions provide a significant benefit to the model in capturing the dynamics of the layer. Here, we again find that increasing the $\Delta t$ of the model can help improve the prediction of temperature and, by reducing drift there, help in predicting velocities as well. We also find a similar relationship with variance in the slow fields as the drift in the Gulf Stream and the African Cape, with larger variances, is larger than in the lower variance regions.

\begin{figure}[h]
    \centering
    \includegraphics[width=1\linewidth]{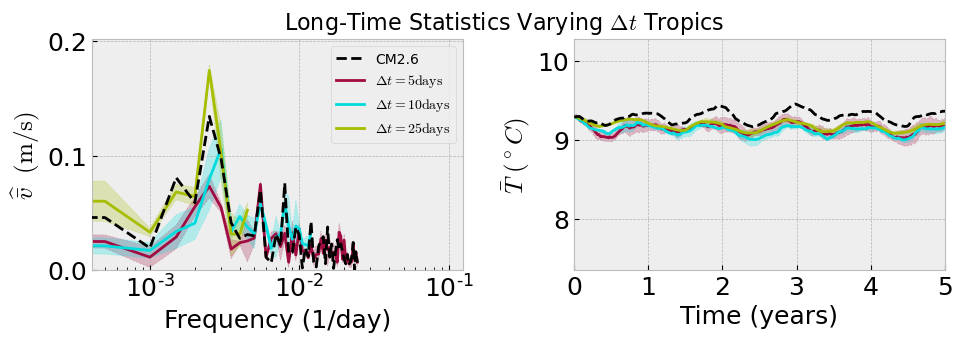}
    \caption{Impact of model time step at a depth of 525 m, showing the skill of each emulator in performing long roll-outs. Left: the time spectra of the mean meridional velocity; Right: mean temperature timeseries. The shading shows the standard deviation across rollouts from 3 networks trained from different initial weights.}
    \label{fig:enter-label}
\end{figure}
\begin{figure}[h]
    \centering
    \includegraphics[width=1\linewidth]{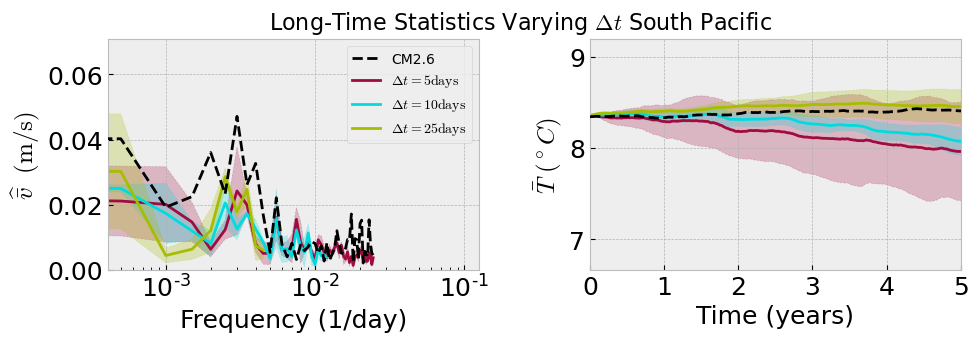}
    \caption{Impact of model time step at a depth of 525 m, showing the skill of each emulator in performing long roll-outs. Left: the time spectra of the mean meridional velocity; Right: mean temperature timeseries. The shading shows the standard deviation across rollouts from 3 networks trained from different initial weights.}
    \label{fig:enter-label}
\end{figure}
\begin{figure}[h]
    \centering
    \includegraphics[width=1\linewidth]{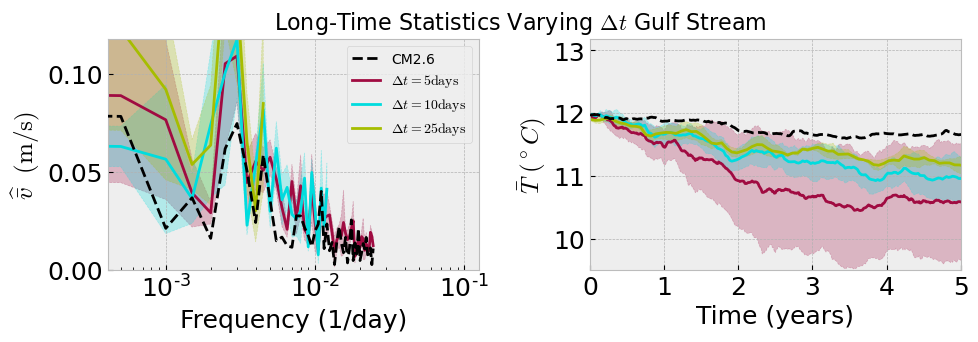}
    \caption{Impact of model time step at a depth of 525 m, showing the skill of each emulator in performing long roll-outs. Left: the time spectra of the mean meridional velocity; Right: mean temperature timeseries. The shading shows the standard deviation across rollouts from 3 networks trained from different initial weights.}
    \label{fig:enter-label}
\end{figure}
\begin{figure}[h]
    \centering
    \includegraphics[width=1\linewidth]{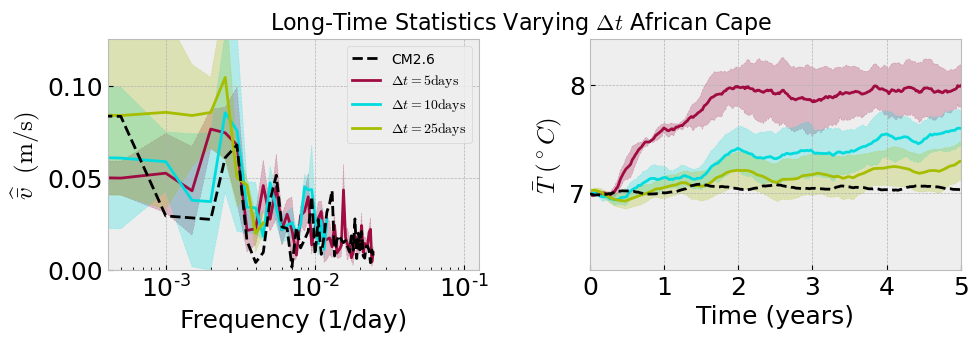}
    \caption{Impact of model time step at a depth of 525 m, showing the skill of each emulator in performing long roll-outs. Left: the time spectra of the mean meridional velocity; Right: mean temperature timeseries. The shading shows the standard deviation across rollouts from 3 networks trained from different initial weights.}
    \label{fig:enter-label}
\end{figure}

\subsection*{Figures Varying Boundary Conditions}
Here we include additional versions of figure \ref{fig:Vary_Boundary_Main} for the remaining regions not shown in the paper. We again omit the results from the African Cape region as all models perform poorly without a longer training interval or time step. We also include some selected snapshots of model prediction from the Gulf Stream and the Tropics. Results from the African Cape and the South Pacific will be shown for models varying $\Delta t$ and $N$. For zonal velocities, we show results at 50 days, and for temperature, we show snapshots 1000 days into a rollout to demonstrate that the model captures the correct phase of the seasonal cycle. Meridional velocity results look similar to zonal velocity and are omitted for brevity. We also include here an additional figure showing the time averaged mean kinetic energy in the gulf stream to help highlight the importance of wind stress terms. With just temperature, the emulator misplaces the jet slightly, turning eastward at too high a  latitude.
\begin{figure}
    \centering
    \includegraphics[width=1\linewidth]{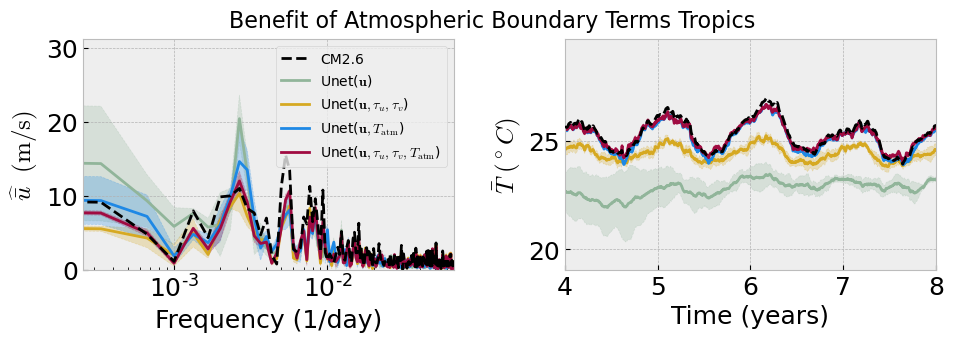}
    \caption{Impact of atmospheric boundary information, showing the skill of each emulator in performing long roll-outs. Left: the time spectra of the mean zonal velocity; Right: mean temperature timeseries. The shading shows the standard deviation across rollouts from 3 networks trained from different initial weights.}
    \label{fig:tropics_boundary}
\end{figure}

\begin{figure}
    \centering
    \includegraphics[width=1\linewidth]{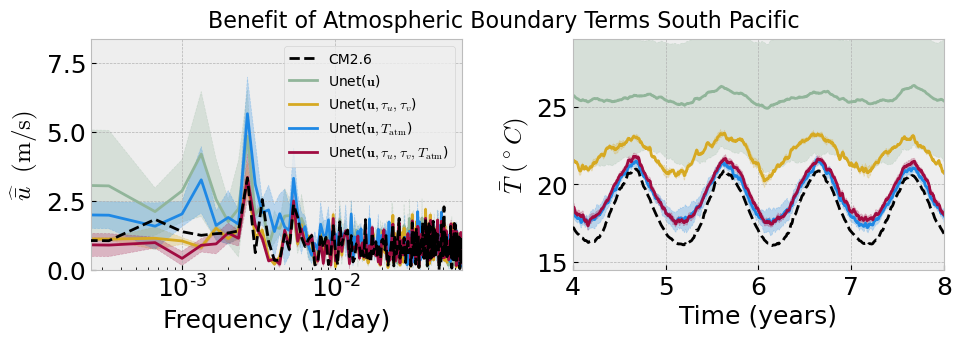}
    \caption{Impact of atmospheric boundary information, showing the skill of each emulator in performing long roll-outs. Left: the time spectra of the mean zonal velocity; Right: mean temperature timeseries. The shading shows the standard deviation across rollouts from 3 networks trained from different initial weights.}
    \label{fig:enter-label}
\end{figure}

\begin{figure}
    \centering
    \includegraphics[width=1\linewidth]{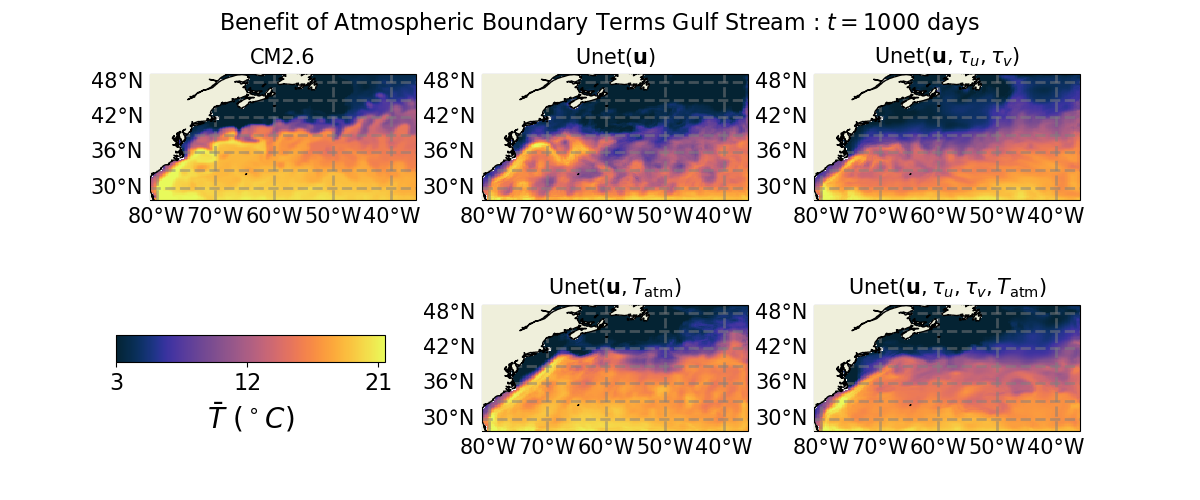}
    \caption{Snapshots of temperature shown for a 1000 days prediction, generated by the emulators with different atmospheric boundary inputs (as in fig.~\ref{fig:Vary_Boundary_Main}.)}
    \label{fig:enter-label}
\end{figure}

\begin{figure}
    \centering
    \includegraphics[width=1\linewidth]{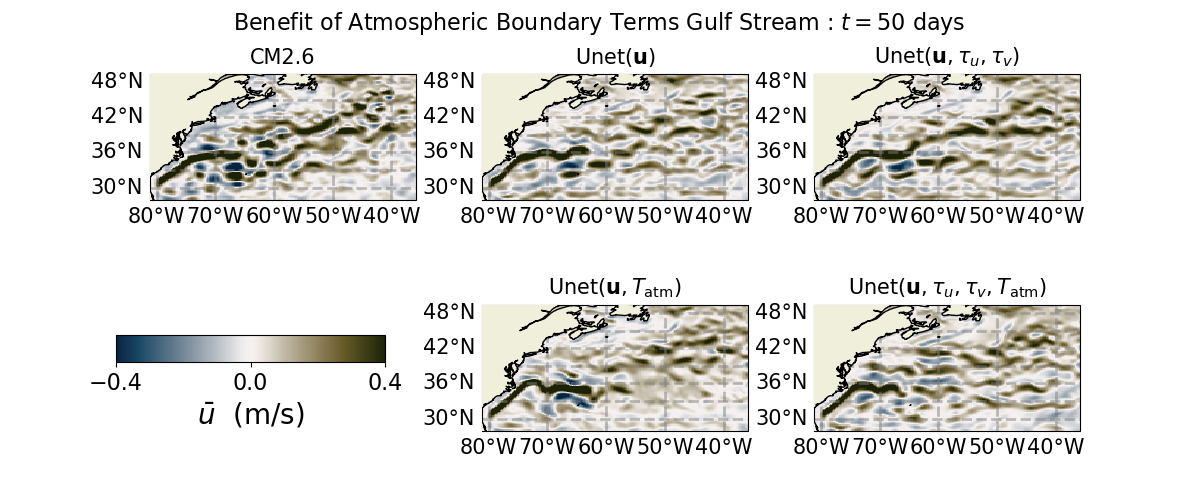}
    \caption{Snapshots of zonal velocity shown for a 50 days prediction, generated by the emulators with different atmospheric boundary inputs (as in fig.~\ref{fig:Vary_Boundary_Main}.)}
    \label{fig:enter-label}
\end{figure}

\begin{figure}
    \centering
    \includegraphics[width=1\linewidth]{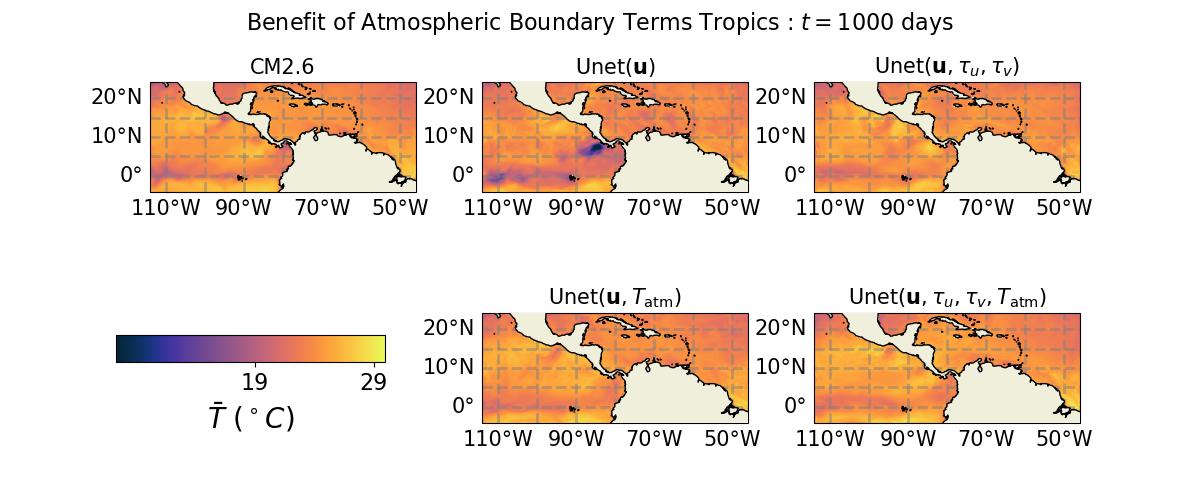}
    \caption{Snapshots of temperature shown for a 1000 days prediction, generated by the emulators with different atmospheric boundary inputs (as in fig.~\ref{fig:tropics_boundary}.)}
    \label{fig:enter-label}
\end{figure}

\begin{figure}
    \centering
    \includegraphics[width=1\linewidth]{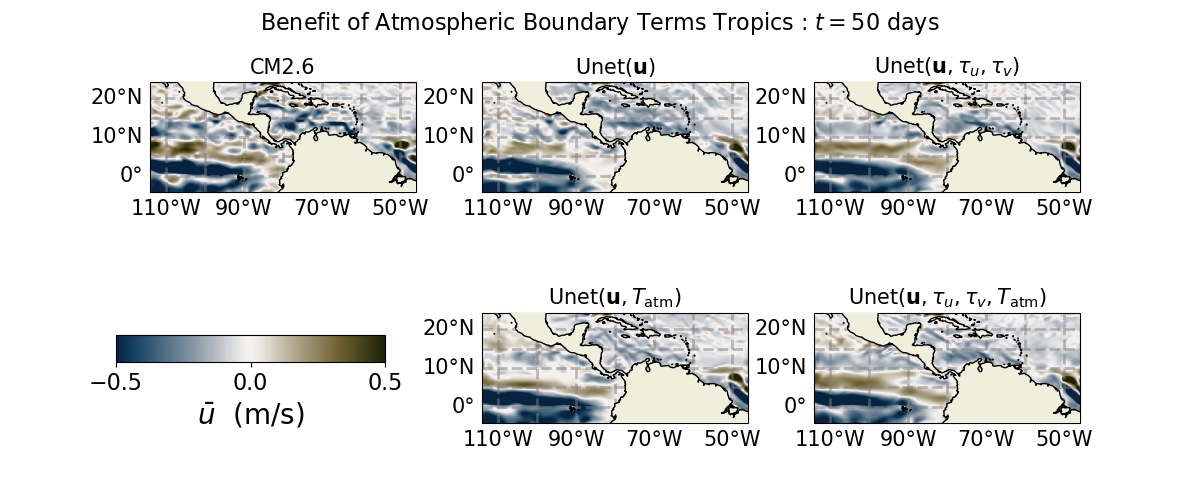}
    \caption{Snapshots of zonal velocity shown for a 50 days prediction, generated by the emulators with different atmospheric boundary inputs (as in fig.~\ref{fig:tropics_boundary}.)}
    \label{fig:enter-label}
\end{figure}

\begin{figure}
    \centering
    \includegraphics[width=1\linewidth]{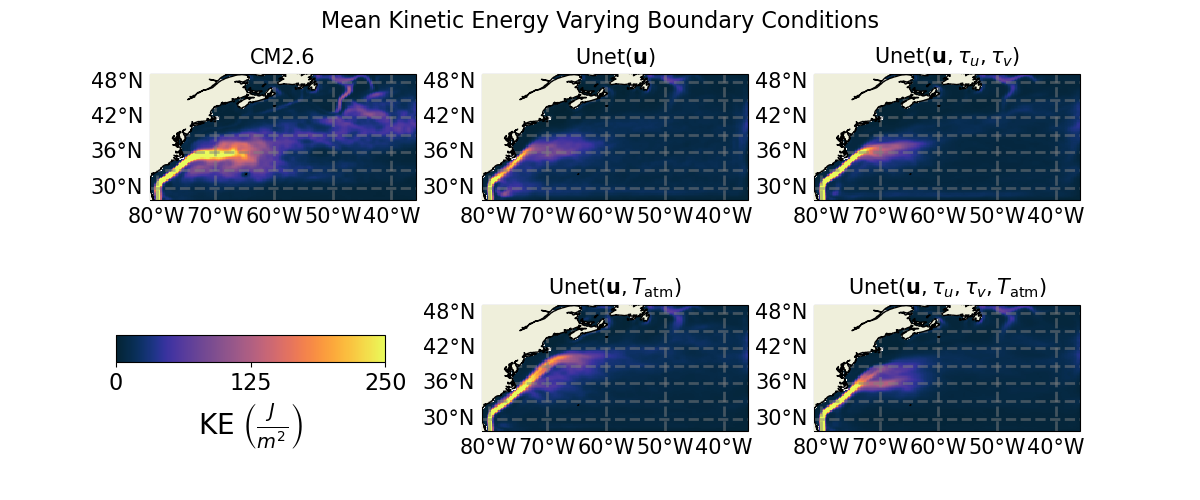}
    \caption{Temporal means of kinetic energy shown averaged over 3000 day emulator rollouts generated by the emulators with different atmospheric boundary inputs (as in fig.~\ref{fig:tropics_boundary}).}
    \label{fig:enter-label}
\end{figure}

\subsection*{Figures Varying the Training Window}
Here we include additional versions of figure \ref{fig:Lag_Main} for the remaining regions not shown in the paper as well as longer statistics for the same set of models. We also include some selected snapshots of model prediction from the  African Cape and the South Pacific. The snapshots from the African Cape region show results with models all matching the training window of the base model. These models match those shown in the bottom row of figure \ref{fig:Lag_Main}. The snapshots from the South Pacific show results with models varying the time step. These models match those shown in the top row of figure \ref{fig:Lag_Main}. We also include tables comparing the ACC and RMSE evaluated at 10 days for emulators in each region.
\begin{figure}[h]
    \centering
    \includegraphics[width=1\linewidth]{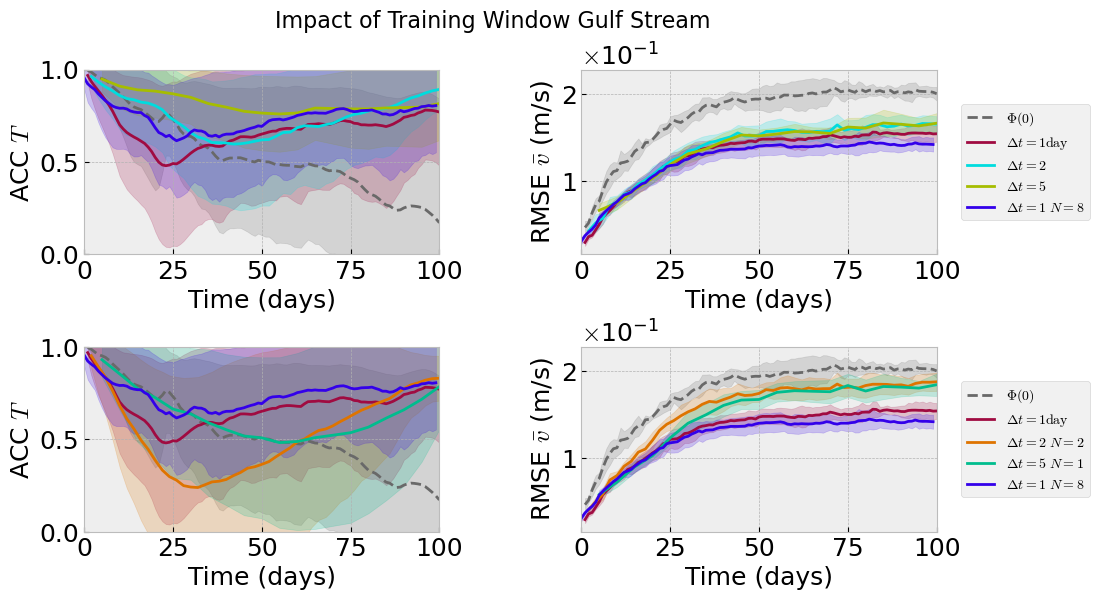}
    \caption{Impact of time step and effective training window on the short-term skill of the network for the Gulf Stream region. ACC for temperature (left) and RMSE for meridional velocity (right). The top panels showcase the benefit of increasing $\Delta t$, and the bottom panels demonstrate the performance loss when decreasing the number of rollouts, $N$. The shading indicates the standard deviation across rollouts from 5 initial conditions and using 3 networks trained from different initial weights. }
    \label{fig:enter-label}
\end{figure}

\begin{figure}[h]
    \centering
    \includegraphics[width=1\linewidth]{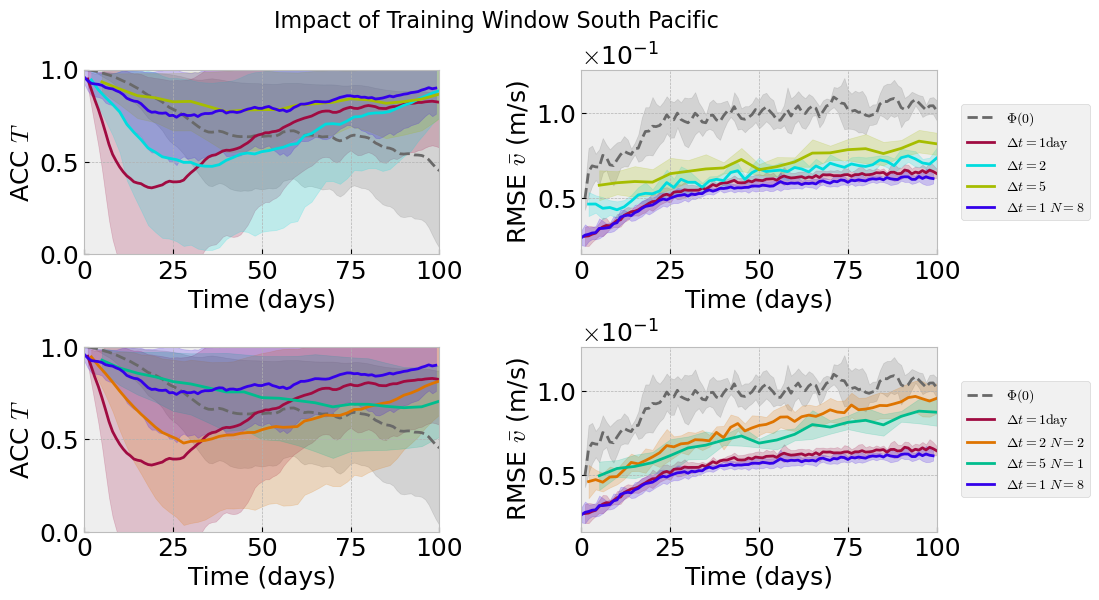}
    \caption{Impact of time step and effective training window on the short-term skill of the network for the South Pacific region. ACC for temperature (left) and RMSE for meridional velocity (right). The top panels showcase the benefit of increasing $\Delta t$, and the bottom panels demonstrate the performance loss when decreasing the number of rollouts, $N$. The shading indicates the standard deviation across rollouts from 5 initial conditions and using 3 networks trained from different initial weights.}
    \label{fig:lag_south_pacific}
\end{figure}

\begin{figure}[h]
    \centering
    \includegraphics[width=1\linewidth]{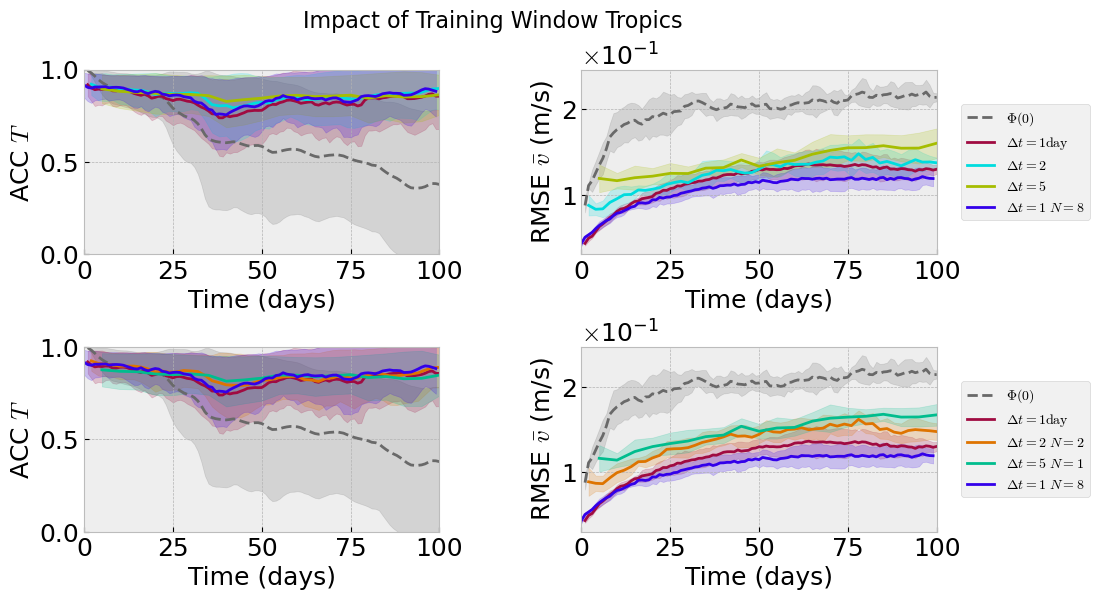}
    \caption{Impact of time step and effective training window on the short-term skill of the network for the Tropics region. ACC for temperature (left) and RMSE for meridional velocity (right). The top panels showcase the benefit of increasing $\Delta t$, and the bottom panels demonstrate the performance loss when decreasing the number of rollouts, $N$. The shading indicates the standard deviation across rollouts from 5 initial conditions and using 3 networks trained from different initial weights.}
    \label{fig:enter-label}
\end{figure}

\begin{figure}[ht]
    \centering
    \includegraphics[width=1\linewidth]{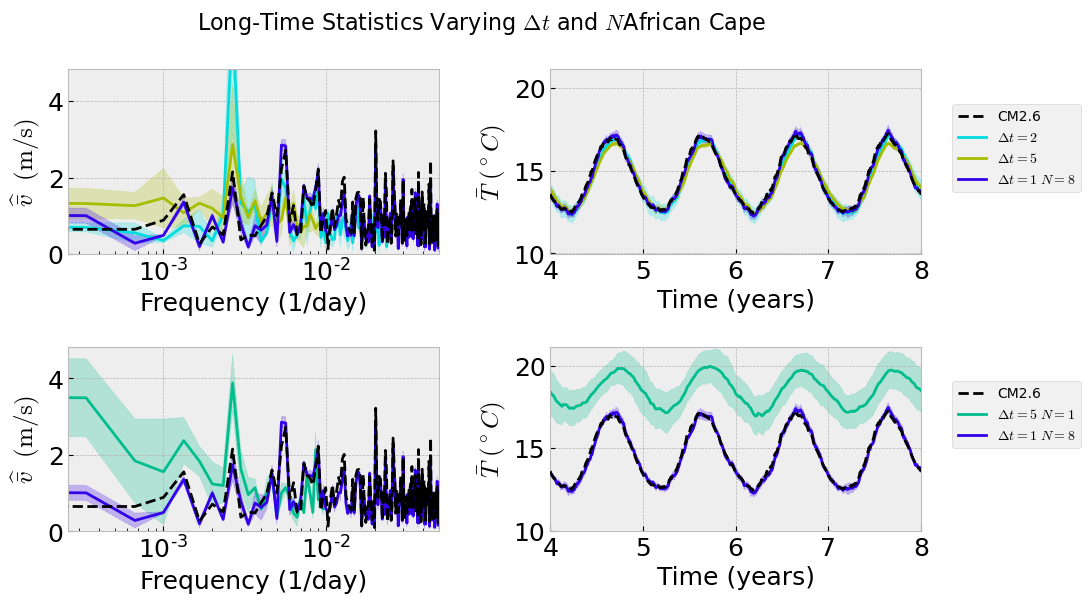}
    \caption{Impact of model time step, showing the skill of each emulator in performing long roll-outs. Left: the time spectra of the mean zonal velocity; Right: mean temperature timeseries. The shading shows the standard deviation across rollouts from 3 networks trained from different initial weights. Lines for $\Delta t =1$ and $\Delta t = 2,~ N = 2$ are omitted since these predictions drift heavily.}
    \label{fig:enter-label}
\end{figure}

\begin{figure}[ht]
    \centering
    \includegraphics[width=1\linewidth]{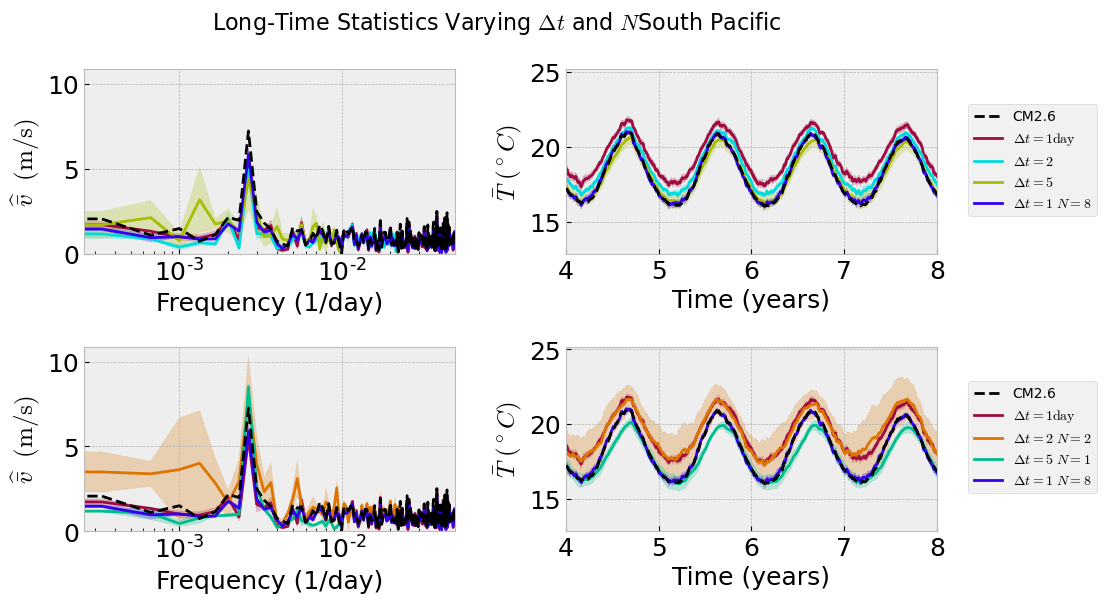}
    \caption{Impact of model time step, showing the skill of each emulator in performing long roll-outs. Left: the time spectra of the mean zonal velocity; Right: mean temperature timeseries. The shading shows the standard deviation across rollouts from 3 networks trained from different initial weights.  }
    \label{fig:enter-label}
\end{figure}

\begin{figure}[ht]
    \centering
    \includegraphics[width=1\linewidth]{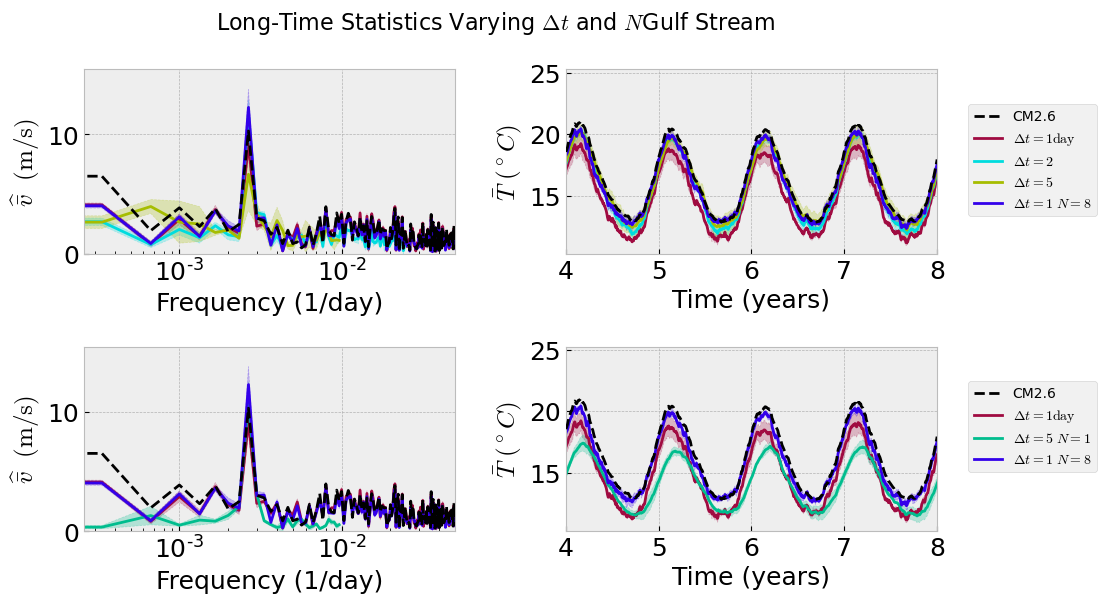}
    \caption{Impact of model time step, showing the skill of each emulator in performing long roll-outs. Left: the time spectra of the mean zonal velocity; Right: mean temperature timeseries. The shading shows the standard deviation across rollouts from 3 networks trained from different initial weights. The line for  $\Delta t = 2,~ N = 2$ is omitted since the emulator becomes unstable.}
    \label{fig:enter-label}
\end{figure}
\begin{figure}[ht]
    \centering
    \includegraphics[width=1\linewidth]{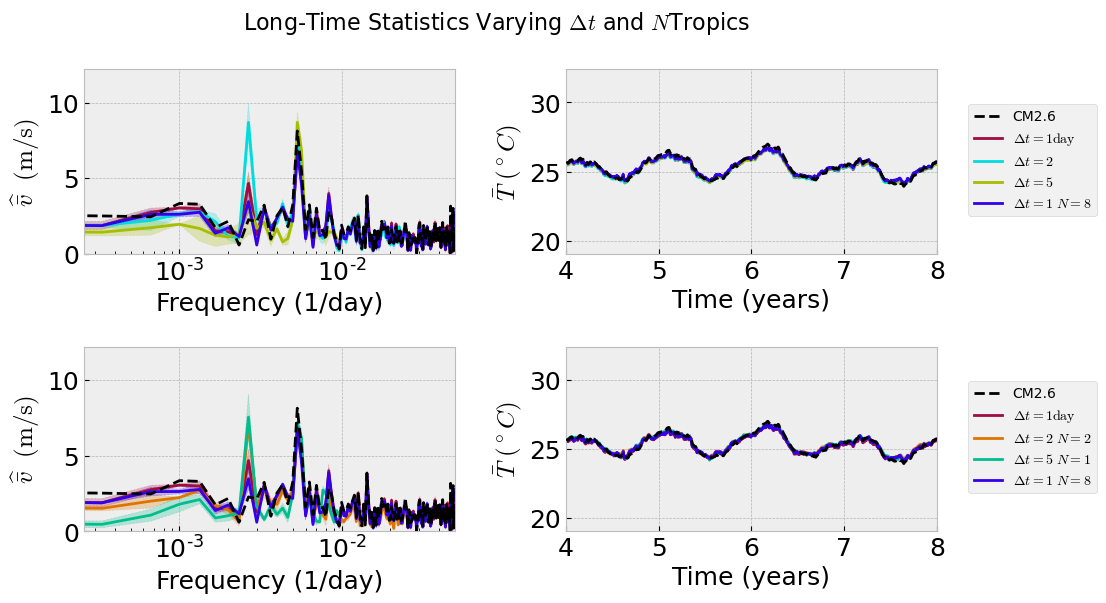}
    \caption{Impact of model time step, showing the skill of each emulator in performing long roll-outs. Left: the time spectra of the mean zonal velocity; Right: mean temperature timeseries. The shading shows the standard deviation across rollouts from 3 networks trained from different initial weights.}
    \label{fig:enter-label}
\end{figure}

\begin{figure}[ht]
    \centering
    \includegraphics[width=1\linewidth]{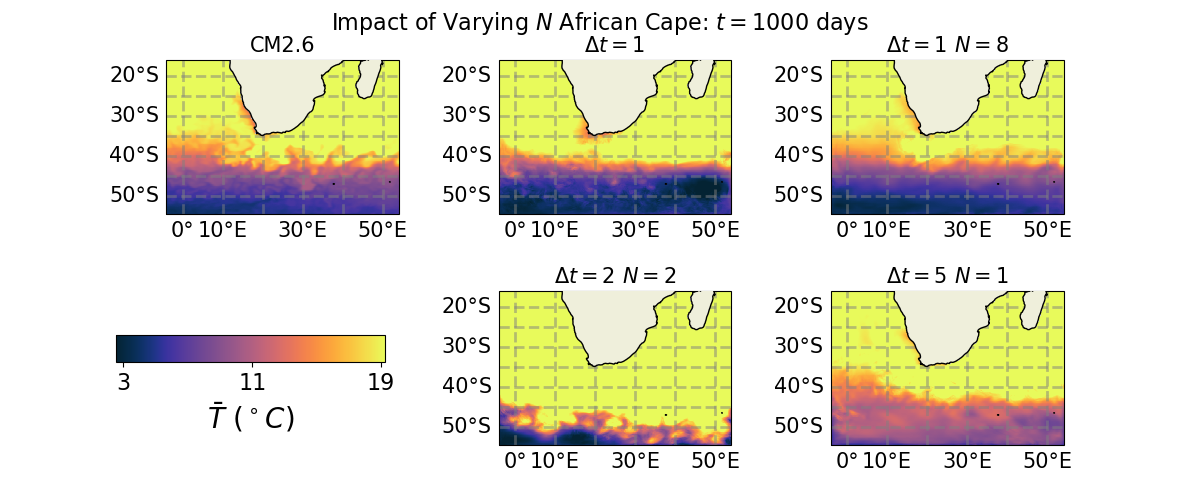}
    \caption{Snapshots of temperature shown for a 1000 days prediction, generated by the emulators with different $\Delta t$ and $N$  (as in fig.~\ref{fig:Lag_Main}.) }
    \label{fig:enter-label}
\end{figure}

\begin{figure}[ht]
    \centering
    \includegraphics[width=1\linewidth]{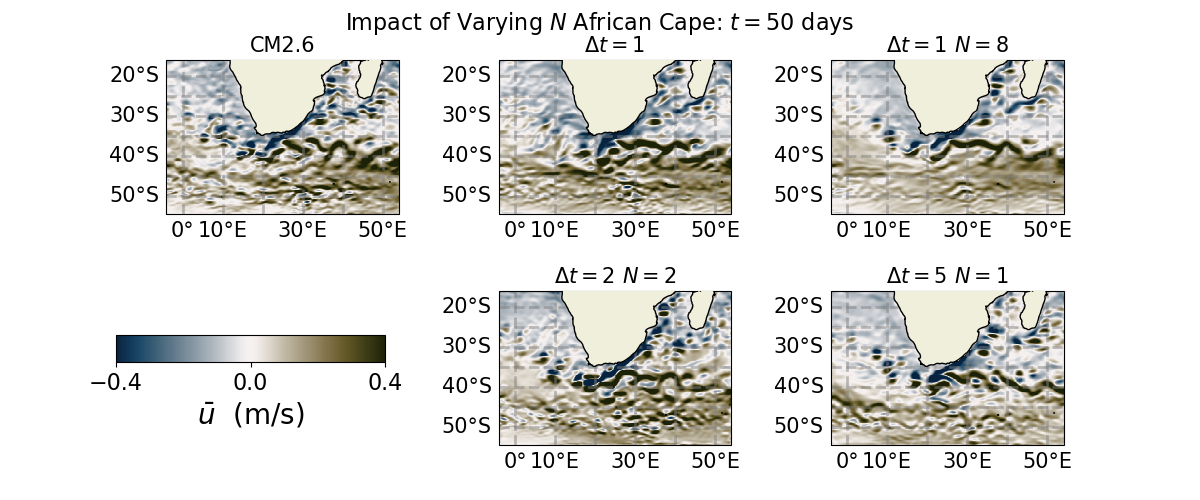}
    \caption{Snapshots of zonal velocity shown for a 50 days prediction, generated by the emulators with different $\Delta t$ and $N$  (as in fig.~\ref{fig:Lag_Main}.) }
    \label{fig:enter-label}
\end{figure}

\begin{figure}[ht]
    \centering
    \includegraphics[width=1\linewidth]{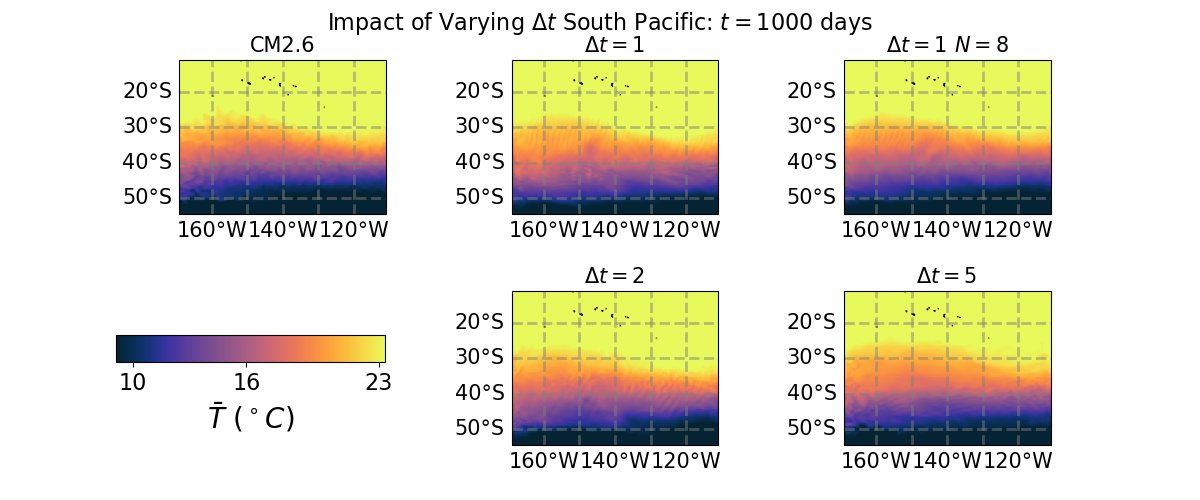}
    \caption{Snapshots of temperature shown for a 1000 days prediction, generated by the emulators with different $\Delta t$ and $N$  (as in fig.~\ref{fig:lag_south_pacific}.) }
    \label{fig:enter-label}
\end{figure}

\begin{figure}[ht]
    \centering
    \includegraphics[width=1\linewidth]{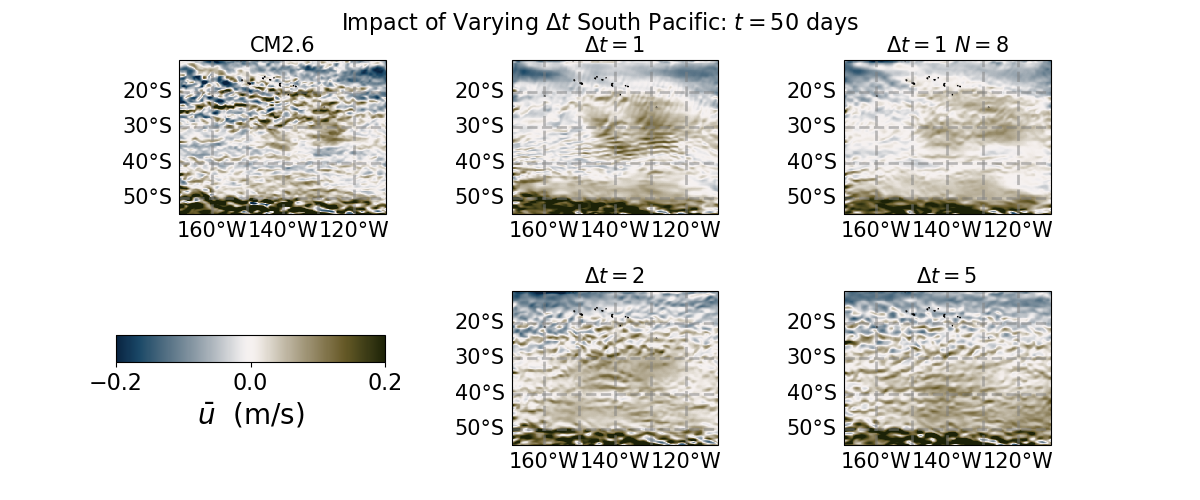}
    \caption{Snapshots of zonal velocity shown for a 50 days prediction, generated by the emulators with different $\Delta t$ and $N$  (as in fig.~\ref{fig:lag_south_pacific}.) }
    \label{fig:enter-label}
\end{figure}

\begin{table}[h]
    \centering
\caption{Table for the all regions showing the RMSE and ACC evaluated at 10 days.}
\label{tab:my_label}
    \begin{tabular}{|c|c|c|c|c|c|c|c|} \hline
 \multicolumn{8}{|c|}{Error Metrics for the South Pacific at $t = 10$ days}\\ \hline
         $\Delta t$&  1&  2&  5&  1&  2&  5
& \\ \hline
         $N$&  4&  4&  4&  8&  2&  1
& $\boldsymbol{\Phi}(0)$\\ \hline
         RMSE u&  0.040&  0.047&  0.066&  0.038&  0.058&  0.063& 0.075
\\ \hline
         RMSE v&  0.037&  0.043&  0.059&  0.035&  0.049&  0.054& 0.073
\\ \hline
         RMSE T&  0.965&  0.596&  0.457&  0.402&  0.704&  0.612& 0.336
\\ \hline
         ACC u&  0.866&  0.802&  0.655&  0.881&  0.708&  0.683& 0.590
\\ \hline
         ACC v&  0.870&  0.813&  0.648&  0.888&  0.764&  0.707& 0.525
\\ \hline
         ACC T&  0.473&  0.771&  0.895&  0.886&  0.760&  0.885& 0.947
\\ \hline
 \multicolumn{8}{|c|}{Error Metrics for the African Cape at $t = 10$ days}\\\hline
 RMSE u& 0.074& 0.078& 0.085& 0.068& 0.094& 0.092&0.137
\\\hline

RMSE v& 0.075& 0.077& 0.084& 0.070& 0.091& 0.089&0.142
\\\hline
 RMSE T& 1.152& 0.809& 0.624& 0.834& 1.203& 0.925&0.602
\\\hline

ACC u& 0.875& 0.864& 0.835& 0.893& 0.813& 0.810&0.598
\\\hline
 ACC v& 0.894& 0.886& 0.866& 0.905& 0.850& 0.854&0.634
\\\hline

ACC T& 0.732& 0.838& 0.906& 0.815& 0.731& 0.794&0.896
\\ \hline
 \multicolumn{8}{|c|}{Error Metrics for the Gulf Stream at $t = 10$ days}\\\hline
 RMSE u& 0.076& 0.080& 0.074& 0.074& 0.084& 0.074
&0.112\\\hline
 RMSE v& 0.075& 0.075& 0.074& 0.073& 0.080& 0.072
&0.112
\\\hline
 RMSE T& 1.461& 1.018& 0.828& 1.165& 1.538& 1.046
&0.822
\\\hline
 ACC u& 0.862& 0.843& 0.864& 0.867& 0.835& 0.864
&0.713
\\\hline
 ACC v& 0.855& 0.855& 0.851& 0.863& 0.844& 0.863
&0.688
\\\hline
 ACC T& 0.701& 0.859& 0.910& 0.795& 0.681& 0.853
&0.191
\\ \hline
 \multicolumn{8}{|c|}{Error Metrics for the Tropics at $t = 10$ days}\\\hline
 RMSE u& 0.088& 0.095& 0.12& 0.081& 0.107& 0.126
&0.161
\\\hline
 RMSE v& 0.08& 0.095& 0.116& 0.075& 0.099& 0.114
&0.176
\\\hline

RMSE T& 1.461& 1.018& 0.828& 1.165& 1.538& 1.046
&0.822
\\\hline
 ACC u& 0.873& 0.859& 0.781& 0.888& 0.81& 0.755
&0.642
\\\hline
 ACC v& 0.85& 0.785& 0.663& 0.868& 0.771& 0.681
&0.329
\\\hline

ACC T& 0.889& 0.896& 0.882& 0.904& 0.895& 0.867
&0.895
\\\hline
    \end{tabular}

\end{table}

\end{document}

%% file: math_commands.tex
\usepackage{amsmath,amsfonts,bm}

\def\eqref#1{equation~\ref{#1}}

\def\1{\bm{1}}

\DeclareMathAlphabet{\mathsfit}{\encodingdefault}{\sfdefault}{m}{sl}
\SetMathAlphabet{\mathsfit}{bold}{\encodingdefault}{\sfdefault}{bx}{n}













%% file: main.bbl
\begin{thebibliography}{14}
\providecommand{\natexlab}[1]{#1}
\providecommand{\url}[1]{\texttt{#1}}
\expandafter\ifx\csname urlstyle\endcsname\relax
  \providecommand{\doi}[1]{doi: #1}\else
  \providecommand{\doi}{doi: \begingroup \urlstyle{rm}\Url}\fi

\bibitem[Beucler et~al.(2023)Beucler, Ebert-Uphoff, Rasp, Pritchard, and
  Gentine]{beucler2023machine}
Tom Beucler, Imme Ebert-Uphoff, Stephan Rasp, Michael Pritchard, and Pierre
  Gentine.
\newblock Machine learning for clouds and climate.
\newblock \emph{Clouds and their Climatic Impacts: Radiation, Circulation, and
  Precipitation}, pp.\  325--345, 2023.

\bibitem[Bi et~al.(2023)Bi, Xie, Zhang, Chen, Gu, and Tian]{bi2023accurate}
Kaifeng Bi, Lingxi Xie, Hengheng Zhang, Xin Chen, Xiaotao Gu, and Qi~Tian.
\newblock Accurate medium-range global weather forecasting with 3d neural
  networks.
\newblock \emph{Nature}, 619\penalty0 (7970):\penalty0 533--538, 2023.

\bibitem[Bire et~al.(2023)Bire, L{\"u}tjens, Azizzadenesheli, Anandkumar, and
  Hill]{bire2023ocean}
Suyash Bire, Bj{\"o}rn L{\"u}tjens, Kamyar Azizzadenesheli, Anima Anandkumar,
  and Christopher~N Hill.
\newblock Ocean emulation with fourier neural operators: Double gyre.
\newblock \emph{Authorea Preprints}, 2023.

\bibitem[Bonev et~al.(2023)Bonev, Kurth, Hundt, Pathak, Baust, Kashinath, and
  Anandkumar]{bonev2023spherical}
Boris Bonev, Thorsten Kurth, Christian Hundt, Jaideep Pathak, Maximilian Baust,
  Karthik Kashinath, and Anima Anandkumar.
\newblock Spherical fourier neural operators: Learning stable dynamics on the
  sphere.
\newblock \emph{arXiv preprint arXiv:2306.03838}, 2023.

\bibitem[Chattopadhyay et~al.(2023)Chattopadhyay, Gray, Wu, Lowe, and
  He]{chattopadhyay2023oceannet}
Ashesh Chattopadhyay, Michael Gray, Tianning Wu, Anna~B Lowe, and Ruoying He.
\newblock Oceannet: A principled neural operator-based digital twin for
  regional oceans.
\newblock \emph{arXiv preprint arXiv:2310.00813}, 2023.

\bibitem[Delworth et~al.(2012)Delworth, Rosati, Anderson, Adcroft, Balaji,
  Benson, Dixon, Griffies, Lee, Pacanowski, et~al.]{delworth2012simulated}
Thomas~L Delworth, Anthony Rosati, Whit Anderson, Alistair~J Adcroft,
  Venkatramani Balaji, Rusty Benson, Keith Dixon, Stephen~M Griffies, Hyun-Chul
  Lee, Ronald~C Pacanowski, et~al.
\newblock Simulated climate and climate change in the gfdl cm2. 5
  high-resolution coupled climate model.
\newblock \emph{Journal of Climate}, 25\penalty0 (8):\penalty0 2755--2781,
  2012.

\bibitem[Kara et~al.(2007)Kara, Wallcraft, Metzger, Hurlburt, and
  Fairall]{kara2007wind}
A~Birol Kara, Alan~J Wallcraft, E~Joseph Metzger, Harley~E Hurlburt, and
  Chris~W Fairall.
\newblock Wind stress drag coefficient over the global ocean.
\newblock \emph{Journal of Climate}, 20\penalty0 (23):\penalty0 5856--5864,
  2007.

\bibitem[Kochkov et~al.(2023)Kochkov, Yuval, Langmore, Norgaard, Smith, Mooers,
  Lottes, Rasp, D{\"u}ben, Kl{\"o}wer, et~al.]{kochkov2023neural}
Dmitrii Kochkov, Janni Yuval, Ian Langmore, Peter Norgaard, Jamie Smith,
  Griffin Mooers, James Lottes, Stephan Rasp, Peter D{\"u}ben, Milan
  Kl{\"o}wer, et~al.
\newblock Neural general circulation models.
\newblock \emph{arXiv preprint arXiv:2311.07222}, 2023.

\bibitem[Loose et~al.(2022)Loose, Abernathey, Grooms, Busecke, Guillaumin,
  Yankovsky, Marques, Steinberg, Ross, Khatri, Bachman, Zanna, and
  Martin]{Loose2022}
Nora Loose, Ryan Abernathey, Ian Grooms, Julius Busecke, Arthur Guillaumin,
  Elizabeth Yankovsky, Gustavo Marques, Jacob Steinberg, Andrew~Slavin Ross,
  Hemant Khatri, Scott Bachman, Laure Zanna, and Paige Martin.
\newblock Gcm-filters: A python package for diffusion-based spatial filtering
  of gridded data.
\newblock \emph{Journal of Open Source Software}, 7\penalty0 (70):\penalty0
  3947, 2022.
\newblock \doi{10.21105/joss.03947}.
\newblock URL \url{https://doi.org/10.21105/joss.03947}.

\bibitem[Marchesiello et~al.(2001)Marchesiello, McWilliams, and
  Shchepetkin]{marchesiello2001open}
Patrick Marchesiello, James~C McWilliams, and Alexander Shchepetkin.
\newblock Open boundary conditions for long-term integration of regional
  oceanic models.
\newblock \emph{Ocean modelling}, 3\penalty0 (1-2):\penalty0 1--20, 2001.

\bibitem[Pathak et~al.(2022)Pathak, Subramanian, Harrington, Raja,
  Chattopadhyay, Mardani, Kurth, Hall, Li, Azizzadenesheli,
  et~al.]{pathak2022fourcastnet}
Jaideep Pathak, Shashank Subramanian, Peter Harrington, Sanjeev Raja, Ashesh
  Chattopadhyay, Morteza Mardani, Thorsten Kurth, David Hall, Zongyi Li, Kamyar
  Azizzadenesheli, et~al.
\newblock Fourcastnet: A global data-driven high-resolution weather model using
  adaptive fourier neural operators.
\newblock \emph{arXiv preprint arXiv:2202.11214}, 2022.

\bibitem[Ronneberger et~al.(2015)Ronneberger, Fischer, and
  Brox]{ronneberger2015u}
Olaf Ronneberger, Philipp Fischer, and Thomas Brox.
\newblock U-net: Convolutional networks for biomedical image segmentation.
\newblock In \emph{Medical Image Computing and Computer-Assisted
  Intervention--MICCAI 2015: 18th International Conference, Munich, Germany,
  October 5-9, 2015, Proceedings, Part III 18}, pp.\  234--241. Springer, 2015.

\bibitem[Watt-Meyer et~al.(2023)Watt-Meyer, Dresdner, McGibbon, Clark, Henn,
  Duncan, Brenowitz, Kashinath, Pritchard, Bonev, et~al.]{watt2023ace}
Oliver Watt-Meyer, Gideon Dresdner, Jeremy McGibbon, Spencer~K Clark, Brian
  Henn, James Duncan, Noah~D Brenowitz, Karthik Kashinath, Michael~S Pritchard,
  Boris Bonev, et~al.
\newblock Ace: A fast, skillful learned global atmospheric model for climate
  prediction.
\newblock \emph{arXiv preprint arXiv:2310.02074}, 2023.

\bibitem[Xiong et~al.(2023)Xiong, Xiang, Wu, Zhou, Sun, Ma, and
  Huang]{xiong2023ai}
Wei Xiong, Yanfei Xiang, Hao Wu, Shuyi Zhou, Yuze Sun, Muyuan Ma, and Xiaomeng
  Huang.
\newblock Ai-goms: Large ai-driven global ocean modeling system.
\newblock \emph{arXiv preprint arXiv:2308.03152}, 2023.

\end{thebibliography}
